\documentclass[prl,showpacs]{revtex4}

\usepackage{graphicx}
\usepackage{amsmath}
\usepackage[bf,small]{caption2}
\usepackage{floatflt}
\usepackage{bm}

\begin{document}

\title{Electron localization in two-dimensional surface-corrugated conductors: \\
manifestation of competing scattering mechanisms}

\author{N. M. Makarov}
\email{makarov@sirio.ifuap.buap.mx}

\affiliation{C.I.D.S., Instituto de Ciencias,
         Universidad Aut\'{o}noma de Puebla, \\
         Priv. 17 Norte No 3417, Col. San Miguel Hueyotlipan,
         Puebla, Pue., 72050, M\'{e}xico}

\author{Yu.V. Tarasov}
\email{yutarasov@ire.kharkov.ua}

\affiliation{Institute for Radiophysics \& Electronics NASU,
 12 Acad. Proskura St., Kharkov 61085, Ukraine}

\date{\today}

%------------------------------------------------------------------
\begin{abstract}
Transport properties of narrow two-dimensional conducting wires in
which the electron scattering is caused by side edges' roughness
have been studied. The method for calculating dynamic
characteristics of such conductors is proposed which is based on
the two-scale representation of the mode wave functions at weak
scattering. With this method, fundamentally different {\it
by-height} and {\it by-slope} scattering mechanisms associated
with edge roughness are discriminated. The results for single-mode
systems, previously obtained by conventional methods, are proven
to correspond to the former mechanism only. Yet the commonly
ignored by-slope scattering is more likely dominant. The electron
extinction lengths relevant to this scattering differ
substantially in functional structure from those pertinent to the
by-height scattering. The transmittance of ultra-quantum wires is
calculated over all range of scattering parameters, from ballistic
to localized transport of quasi-particles. The obtained dependence
of scattering lengths on the disorder parameters is valid
qualitatively for arbitrary inter-correlation of the boundaries'
defects.
\end{abstract}
%-------------------------------------------------------------------

\pacs{72.10.-d; 72.15.Rn; 73.20.Fz; 73.23.-b}

\maketitle

%------------------------------------------------------------------
\section{Introduction}
%------------------------------------------------------------------

The influence of inhomogeneities and defects of different types on
transmission properties of waveguides for both quantum and
classical waves is a subject of long-term intensive research. Most
of theoretical studies have been restricted to flat surfaces. In
practice, however, side boundaries of mesoscopic conductors are
either inherently rough (due to growth, fracture, etc.), or
artificially patterned (e.g. due to lithographic preparation). One
way or the other, surface-corrugated conducting systems find
widespread applications in technology and material science
(quantum conductors, optical fibers, etc.), and, hence, it is
often necessary to know how the shape of the confining surface
affects the charge or classical wave transport in different
systems.

Of peculiar interest for contemporary microelectronics, from both
applied and theoretical viewpoints, are two-dimensional (2D)
conductors of mesoscopic size. Nowadays there exists widely
established opinion (see, e.g.,
Refs.~\cite{TrAsh,MakYur,KL,TakFer,Kun,KozKrokh,MeyStep,%
MakMorYam,BratRash96,MakTar,Sanchez98,Sanchez99} and references
therein) that dynamic properties of pure-in-bulk quantum wires, in
particular two-dimensional, are largely determined by scattering
the electrons from randomly rough side boundaries of the
conductor. This scattering mechanism is proven to be responsible
for both relaxation processes in multi-mode conductors
\cite{MeyStep,MakMorYam,BratRash96} and for non-dissipative
(Anderson) localization of conduction electrons in narrow
single-mode wires \cite{MakYur,MakTar}.

In studying the electron transport in 2D surface-corrugated
systems two main problems are especially highlighted. One of them
is relevant to adequate description of the electron scattering
from statistically rough surfaces. The other, dynamic, problem is
pertinent to proper consideration of the interference of multiply
scattered quantum waves, which is essential for describing within
the framework of a perturbation theory the effects resulting from
non-dissipative localization of the electron states.

To resolve the first problem and make an adequate comparison with
experimental data one needs a theory which relates transport
properties of the conduction electrons to the shape of bounding
surfaces of the conductor. Since in random-inhomogeneous 2D wires
fluctuations of both of the side boundaries can be considered as
either mutually independent or correlative, subject to the
preparation technology, it is important to trace the relation
between the kinetic quantities and the statistics of boundary
irregularities, as well as mutual correlation of the opposite
boundaries.

In the previous paper \cite{MakTar} we have studied the case of a
single-mode 2D conductor with statistically identical rough side
boundaries. The particular model considered in Ref.~\cite{MakTar}
of the wire with completely correlated boundaries (CCB) is
equivalent to the deterministic waveguide system of constant width
whose inhomogeneities consist in solely waveguide bends. It was
shown that the electron dynamics in such conducting strips is
governed by quite a different Hamiltonian than that pertinent to
the seemingly more general model of 2D conductor with one boundary
which is rough and the other being ideally flat \cite{MakYur}. A
substantial distinction of the Hamiltonians in Ref.~\cite{MakYur}
and Ref.~\cite{MakTar} have resulted in qualitative functional
distinction of the obtained scattering lengths.

The results of Refs.~\cite{MakYur} and \cite{MakTar} have provided
the grounds to propose that in those works the quantum wave
scattering can be assumed to be associated with fundamentally
different physical factors, namely with deviation of the
boundaries from their ``ideal'' shape in Ref.~\cite{MakYur} and
fluctuation of their slopes in Ref.~\cite{MakTar}. Provided the
supposition is true, when studying particle or classical wave
transport in surface-corrugated waveguide systems one has to
distinguish between two different non-interfering scattering
mechanisms which we call by-height (BH) and by-slope (BS)
scattering. However, it should be noted that quantum wave
scattering was analyzed in Refs.~\cite{MakYur} and \cite{MakTar}
for different waveguide geometries and with the use of
substantially different methods. Therefore, the conjecture stated
in Ref.~\cite{MakTar} about the relative significance of these
scattering mechanisms needs to be additionally substantiated.

To avoid possible misunderstanding and support our idea of
different scattering mechanisms pertaining to the imperfect
boundaries of guiding systems, we examine the waveguide
(conductor) geometry admitting of both BH and BS scattering
simultaneously. We consider a 2D conducting strip with
statistically symmetri\-cal rough boundaries (the abbreviation SSB
will be used for such a strip, in contrast to the CCB strip
considered in Ref.~\cite{MakTar}), which is physically equivalent
to the waveguide with a straight central line (guiding axis) and a
randomly fluctuating width. It will be shown below that the
suggested model of the waveguide corresponds qualitatively to a 2D
wire with opposite side boundaries whose inter-correlation can be
thought of as arbitrary. For the SSB model, as well as for the
arbitrary 2D waveguide, the presence is typical of both BH and BS
scattering mechanisms. It is noteworthy that these mechanisms
compete with one another, depending on the roughness statistical
parameters, even in the simplest case of boundary asperities being
small in height and rather smooth.

Technically, specifying the scattering mechanisms is most
straightforward by reducing the problem of the electron scattering
from complicated boundaries of the conductor to the appropriate
``bulk'' problem specified by a complex Hamiltonian but simple
boundary conditions. Analysis of the problem thus formulated can
often be found much easier than of the problem with complicated
boundary conditions. In some cases it can even be performed
non-perturbatively. In this work, using such an approach, we
managed to reasonably discriminate between BH and BS scattering
mechanisms and analyze their competition in the electron-surface
scattering. Besides, solving the ``surface'' problem in ``bulk''
formulation enabled us to carefully trace such a fine
\emph{spectral} effect as the Anderson localization of current
carriers.

Note that the method for solving the problems of the by-surface
scattering through reduction to the Hamiltonian form is not quite
original. It was employed, in particular, in Refs.~\cite{MeyStep}
and \cite{BratRash96} where the coordinate transformation was
used, which smooths out the rough surface to the flat one. In this
paper, analogous reduction of a ``surface'' problem to the
Hamiltonian formulation is made by merely going over to the
\emph{local mode representation}. This approach ensures the
optimal choice of trial quantum states which serve as a basis for
the perturbation theory. In our view, it is essential that the
mode states \emph{a priori} contain the information on lateral
confinement of the system under consideration and, therefore, are
more adjusted to perturbative treatment of transport problems in
waveguide-like systems than the widely used isotropic plane-wave
basis.

%--------------------------------------------------------------------
\section{Statement of the problem: choosing statistical model}
%--------------------------------------------------------------------

We consider a conducting strip of average width $D$ with the
non-uniform stretch of length $L$ (shaded region in Fig.~1)
occupying in $(x,y)$ plane the area restricted by the inequalities
\begin{eqnarray}
-L/2\leq &x&\leq L/2\ ,
\nonumber \\[-.5\baselineskip]
\label{strip}\\[-.5\baselineskip]
-D/2+\xi_1(x)\leq &y&\leq D/2+\xi_2(x) \ . \nonumber
\end{eqnarray}
%%%%%
%
\begin{figure}[tbh]
\centering
\scalebox{.7}[.7]{\includegraphics[bb=44 567 566
777]{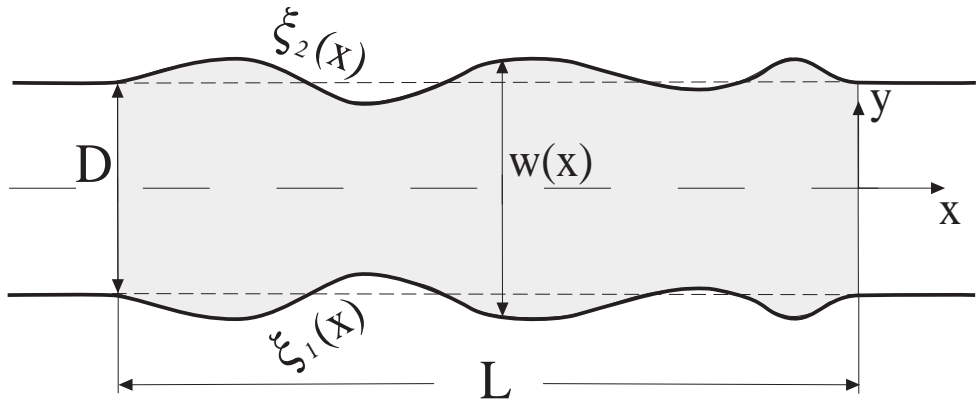}} \caption{$\ $ 2D electron waveguide with rough
side boundaries pertaining to the SSB class.} \label{fig1}
\end{figure}
Functions $\xi_{1,2}(x)$ describing the side boundaries' roughness
are assumed to be random processes with zero mean values. It turns
out to be more convenient for subsequent analysis to introduce,
instead of $\xi_{1,2}(x)$, two different random functions, viz.
\begin{eqnarray}
w(x) = D + \Delta\xi(x)
& \qquad \mbox{and} \qquad &
\xi_c(x) = [\xi_1(x) + \xi_2(x)]/2 \ ,
\label{Width}
\end{eqnarray}
%%%%%
where $\Delta\xi(x) = \xi_2(x) - \xi_1(x)$. Function $w(x)$ is
nothing but the fluctuating width of the strip whereas $\xi_c(x)$
describes the $y$ coordinate of the randomly fluctuating central
(symmetry) line of the 2D waveguide. In terms of the functions
(\ref{Width}), the disordered region (\ref{strip}) is represented
by inequalities
\begin{eqnarray}
-L/2\leq &x&\leq L/2 \ ,
\nonumber\\[-.5\baselineskip]\label{strip_new}\\[-.5\baselineskip]
-w(x)/2+\xi_c(x)\leq &y&\leq w(x)/2+\xi_c(x) \ . \nonumber
\end{eqnarray}
%%%%%
Reformulation of the problem in terms of new characteristic
functions is advantageous from the mathematical point of view.
Besides, it gives a clear idea of the part played by different
physical factors in quantum wave scattering in surface-corrugated
conductors.

Bearing in mind the statistical nature of the problem under
consideration it is necessary to specify correlation properties of
new random variables (\ref{Width}) in conformity with the
conductor geometry chosen. The mean values of the functions $w(x)$
and $\xi_c(x)$ are naturally $\langle w(x)\rangle =D$ and $\langle
\xi_c(x)\rangle =0$, the angular brackets stand for statistical
averaging over realizations of $\xi_{1,2}(x)$. Unlike the plain
averages, the binary correlators of functions (\ref{Width}) are
not so uniquely predetermined but depend significantly on the
intercorrelation between the side boundaries of the conducting
strip. Here we consider the correlation model in whose frames both
of the boundaries are regarded statistically identical in that the
functions $\xi_{1,2}(x)$ obey the equalities
\begin{equation}
\langle \xi_{i}(x) \rangle = 0 \qquad \mbox{and} \qquad \langle
\xi_{i}(x)\xi_{i}(x') \rangle = \sigma^2 {\mathcal W}(x-x') \ ,
\qquad i=1,2 . \label{xi-corrCCB}
\end{equation}
%%%%%
Here $\sigma$ is the rms height of the asperities, which is
thought of to be identical for both of the strip edges, ${\mathcal
W}(x)$ is the correlation coefficient specified by the unit
maximal value and the correlation radius $r_c$.

As for the intercorrelation of the opposite-boundary roughness,
two marginal options seem to be distinct among others for their
particular symmetries. One of them, abbreviated in
Ref.~\cite{MakTar} as CCB model, in terms of statistical
formulation implies the fulfilment of correlation equality
$\langle\xi_{i}(x)\xi_{k}(x')\rangle =\sigma^2{\mathcal W}(x-x')$
at $i\neq k $; it is physically equivalent to a 2D waveguide of
constant width, whose inhomogeneities are in the form of random
bends.

In this work, we make use of another particular model in which the
displacements of the side boundaries comply with the first of
relations (\ref{xi-corrCCB}) whereas for the opposite boundaries
the correlation equality holds
\begin{equation}
\langle\xi_{i}(x)\xi_{k}(x')\rangle =-\sigma^2{\mathcal
W}(x-x') \ , \qquad i\neq k. \label{xi-corrSSB}
\end{equation}
%%%%%
It is easy to make sure that equations (\ref{xi-corrCCB}) and
(\ref{xi-corrSSB}) are equivalent to the following correlations,
%%%
\begin{subequations}
\label{CW}
\begin{eqnarray}
\langle \xi_c(x)w(x') \rangle = 0 \ ,
\label{Center-Width}
\\[6pt]
\langle \xi_c(x)\xi_c(x') \rangle = 0 \ .
\label{Center-Center}
\end{eqnarray}
\end{subequations}
%%%%%
The equality (\ref{Center-Width}) implies that the functions
$\xi_c(x)$ and $w(x)$ can be thought of, within the correlation
approximation, as statistically independent random processes. The
second equality, (\ref{Center-Center}), within the same
approximation is consistent with deterministic relation
$\xi_1(x)=-\xi_2(x)$, that holds true in the case of the electron
waveguide with side boundaries fluctuating symmetrically about the
straight central line, as is shown in Fig.~1 (the SSB model).

From the linear response theory\cite{kubo57}, the dimensionless
conductance $g(L)$ (in units of $e^2/\pi\hbar$) at $T=0$ is
represented by the expression
\begin{equation}
g(L)=-\frac{4}{L^2}\int\!\!\!\!\int d\bm{r}\,d\bm{r}'
\frac{\partial G(\bm{r},\bm{r}')}{\partial x} \frac{\partial
G^*(\bm{r},\bm{r}')}{\partial x'} \ . \label{Kubo_cond}
\end{equation}
%%%%%
Here $G(\bm{r},\bm{r}')$ is the retarded one-electron Green
function; integration with respect to $\bm{r}=(x,y)$ runs over the
area (\ref{strip_new}) occupied by the irregular part of the wire.
Within the isotropic Fermi liquid model, the function
$G(\bm{r},\bm{r}')$ is governed by the equation
\begin{equation}
\left( \Delta+k_F^2+i0\right)G(\bm{r},\bm{r}')=
\delta(\bm{r}-\bm{r}') \ , \label{StartEq}
\end{equation}
%%%%%
where $\Delta$ is the two-dimensional Laplace operator, $k_F$ is
the electron Fermi wave number. As to the boundary conditions to
equation (\ref{StartEq}), in the direction of current ($x$) the
conductor will be regarded as open whereas in the transverse
direction ($y$) the zero (Dirichlet) conditions will be used. Note
that formula (\ref{Kubo_cond}) is valid asymptotically under
conditions of {\it weak scattering} (WS)
\cite{abrikryzh78,efetov83}, which we assume in this paper (see
Eq.~(\ref{weak-scatt}) below).

Inhomogeneity of side boundaries of a waveguide system can be
taken into account in a number of ways. The most commonly used
method reduces to linearization of the exact boundary conditions
in small fluctuations of bounding
surfaces\cite{BassFuks,McGurnMaradud84}. However, such an approach
can result, in some cases, in significant underestimation of the
inter-mode scattering which often exerts primary control over the
dynamics of waves within bounded regions. It will be shown below
that the entanglement of waveguide modes is governed by
fluctuation of the confining surfaces in \emph{slope} rather than
in their \emph{displacement} from the specified ideal shape.
Meanwhile, if one linearizes the boundary conditions, it becomes
precisely the \emph{height} of the surface roughness that serves
as a determining factor for wave scattering. Therefore,
entanglement of the modes can be, either to a large extent or even
entirely, lost when expanding the boundary conditions in small
heights of the surface roughness.

The inter-mode scattering can be taken into account using the
methods of Refs.~\cite{Voronovich94,Voronovich96,DietzeDarling96}
or with the method of ``smoothing'' coordinate transformation,
used in Refs.~\cite{MeyStep,BratRash96,MakTar}. In the latter
papers, the dynamic problem pertinent to the system with
corrugated boundaries is reduced without any approximation to the
analogous problem related to the system with ideal boundaries,
though governed by a more complex Hamiltonian. This method is
advantageous in that it enables one to analyze the dynamics of
quasi-particles without resorting to the concept of
``adiabaticity'' of the confining potential
\cite{GlazLesKhmelShekh88,GlazJons91}.

Technically, the procedure of ``smoothing'' rough boundaries of
the conductor is most convenient to perform without applying
explicitly the coordinate transformation, which, in addition, is
quite non-linear in general case. The same result can be obtained
by going to the local (in lengthwise coordinate $x$) mode
representation in Eqs.~(\ref{Kubo_cond}) and (\ref{StartEq}), i.e.
by performing Fourier transformation in the coordinate $y$ using
the complete set of ``transverse'' eigenfunctions of the Laplace
operator which are consistent with boundary conditions prescribed
at the true walls of the wire, $y=\pm w(x)+\xi_c(x)$. In our case
(Dirichlet conditions) we choose, for definiteness, the functions
\begin{equation}
S_n(y|x) = \left[ \frac{2}{w(x)} \right]^{1/2}
\sin \left[ \left(\frac{y-\xi_c(x)}{w(x)} +
\frac{1}{2} \right) \pi n \right] \ , \qquad
n\in\aleph \ .
\label{Sn_set}
\end{equation}
%%%%%
By substituting $G(\bm{r},\bm{r}')$ in the form of a double
Fourier series,
\begin{equation}
G(\bm{r},\bm{r}') = \sum_{n,n'=1}^{\infty}
S_n(y|x)G_{nn'}(x,x')S_{n'}(y'|x') \ , \label{ModeRep}
\end{equation}
%%%%%
into equation (\ref{StartEq}) we arrive at the following set of
equations for the coefficients $G_{nn'}$ of that series which we
call hereafter the mode Green functions,
\begin{eqnarray}
\left\{ \frac{\partial^2}{\partial x^2} \right. + \left. k_F^2 +i0
- \left( \frac{\pi n}{w(x)} \right)^2 \left[ 1 + {{\xi'_c}^2}(x)
\right]  - \left( \frac{w'(x)}{2w(x)} \right)^2 \left[ 1 +
\frac{(\pi n)^2}{3} \right] \right\}&& G_{nn'}(x,x') \nonumber
\\* - \frac{4}{w(x)} \sum_{m=1\atop (m\neq n)}^{\infty} A_{nm}
\left[ {\hat{\mathcal U}}_{\xi}(x) - C_{nm}
\frac{\xi'_c(x)w'(x)}{w(x)} \right] && G_{mn'}(x,x') \nonumber
\\* + \frac{2}{w(x)} \sum_{m=1\atop (m\neq n)}^{\infty} B_{nm}
\left[ {\hat{\mathcal U}}_{w}(x) - C_{nm} \frac{{w'}^2(x)}{w(x)}
\right] && G_{mn'}(x,x') = \delta_{nn'}\delta(x-x') \
.\qquad\qquad \label{ModEq-2}
\end{eqnarray}
%%%%%
The numerical coefficients in equation (\ref{ModEq-2}) have the
form
\begin{eqnarray}
&& A_{nm} = \frac{nm}{n^2-m^2} \sin^2\left[ \frac{\pi}{2}(n-m)
\right] \ , \nonumber \\
&& B_{nm} = \frac{nm}{n^2-m^2}
\cos^2\left[ \frac{\pi}{2}(n-m) \right] \ , \label{Anm}\\
&&C_{nm}=\frac{3n^2+m^2}{n^2-m^2} \nonumber \ ,
\end{eqnarray}
%%%%%
${\hat{\mathcal U}}_{\xi,w}(x)$ are the differential operators of
the type
\begin{subequations}
\label{UxiUd}
\begin{eqnarray}
&&{\hat{\mathcal U}}_{\xi}(x) = \xi'_c(x)\frac{\partial}{\partial
x} + \frac{\partial}{\partial x}\xi'_c(x) \ , \label{Uxi}\\ &&
{\hat{\mathcal U}}_{w}(x) = w'(x)\frac{\partial}{\partial x} +
\frac{\partial}{\partial x}w'(x) \ . \label{Ud}
\end{eqnarray}
\end{subequations}
%%%%%

The conductance expression (\ref{Kubo_cond}), on substituting
there Green function (\ref{ModeRep}) and integrating over
coordinate $y$, is reduced to the form
\begin{eqnarray}
g(L) = - \frac{4}{L^2} \int_{-L/2}^{L/2}\!\!dx
\int_{-L/2}^{L/2}\!\!dx'\sum_{n,n'=1}^{N_c} && \left[
\frac{\partial G_{nn'}(x,x')}{\partial x} +\sum_{m=1\atop (m\neq
n)}^{\infty}\Phi_{nm}(x)G_{mn'}(x,x')\right] \nonumber \\*
&&\times\left[ \frac{\partial G_{nn'}^*(x,x')}{\partial x'}
-\sum_{m'=1\atop (m'\neq
n')}^{\infty}G_{nm'}^*(x,x')\Phi_{m'n'}(x') \right] \ .
\label{Cond-mode}
\end{eqnarray}
%%%%%
Here $N_c=[kD/\pi]$ is the number of ``open conducting channels'',
that is extended waveguide modes. The coefficient matrix
$\{\Phi_{nm}\}$ in (\ref{Cond-mode}) is composed of the elements
\begin{equation}
\Phi_{nm}(x) = 2B_{nm}\frac{w'(x)}{w(x)}(1-\delta_{nm})
-4A_{nm}\frac{{\xi_c}'(x)}{w(x)} \ ,
\label{Phi_nm(x)}
\end{equation}
whence it is evident, subject to Eq.~(\ref{Anm}), that there is no
diagonal terms, $\Phi_{nn}(x) \equiv 0$.

We emphasize that the set of mode equations
(\ref{ModEq-2})--(\ref{Phi_nm(x)}) is applicable for any
correlation model of rough boundaries. Equation (\ref{ModEq-2})
for the mode Green functions contains the effective
electron-surface scattering potential which consists of two
(substantially different in their physical meaning) types of
terms. The first term, $[\pi n/w(x)]^2$, is determined by random
local displacement of the waveguide boundaries and, therefore, is
responsible for the by-height (BH) scattering. All the other terms
contain the gradients either of the wire width, $w'(x)$, or of the
$y$-coordinate of the symmetry line, ${\xi'}_c(x)$. Consequently,
they describe the by-slope (BS) electron scattering. Note that the
structure of the BS part of the scattering operator is not highly
sensitive to the difference between two specific models of the
waveguide, when either $w'(x)\equiv 0$ or ${\xi'}_c(x)\equiv 0$.
At the same time, depending on these models, the role of the BH
term in the electron scattering changes drastically. Indeed, if
$w'(x)\equiv 0$ (i.e., $w(x)=const$) the electron scattering
arises only due to fluctuations of the roughness slopes that are
described by the BS terms with ${\xi'}_c(x)$. Otherwise, when
${\xi'}_c(x)\equiv 0$ both the BS and BH scattering potentials are
contributing. The case of $w'(x)\equiv 0$ corresponds to the CCB
model of the waveguide \cite{MakTar} (constant width, bend-type
inhomogeneities) whereas the case ${\xi'}_c(x)\equiv 0$ is
physically equivalent to the SSB model with the correlation
properties (\ref{xi-corrCCB})--(\ref{CW}) (``straight'' waveguide
with fluctuating width). So, we can conclude that the SSB model is
qualitatively similar to the general one with arbitrary
correlation properties of the lateral boundaries. On the other
hand, employing this model allows one to avoid excessively
cumbersome calculations without sacrificing the quality of the
results.

%--------------------------------------------
\section{Reduction to one-dimensional dynamic problem}
%--------------------------------------------

In studying the electron transport in bounded systems, one of the
basic characteristics is the number of conducting channels or,
what is the same, the extended waveguide modes. However, the
number of eigen-modes cannot in all cases serve as a good quantum
parameter if one addresses the waveguide with a variable
cross-section. Nevertheless, formulation of the investigated
problem in mode representation is advantageous, since in this
approach the mode structure of the conductor can always be fixed,
$N_c=const$, while boundary displacement can be considered as a
source of perturbation of the mode Hamiltonian in
Eq.~(\ref{ModEq-2}).

In this work, bearing in mind the perspective of Anderson
localization of the current carriers owing to such ``edge
perturbations'', we consider a \emph{single-mode} 2D strip with
the average width $D$ confined within the interval
\begin{equation}
\pi/k_F< D <2\pi/k_F \ .
\label{single-mode}
\end{equation}
For the reasons being discussed below (see also the Appendix),
fluctuations of the conductor width will be regarded small as
compared to its average value.

Under restrictions (\ref{single-mode}), the only element of the
mode Green function matrix $\{G_{nn'}\}$ whose contribution to the
conductance (\ref{Cond-mode}) at weak scattering (see
Eq.~(\ref{weak-scatt})) is not parametrically small is the {\em
intra-mode} propagator $G_{11}(x,x')$. Before deriving the closed
equation for this function, it is worth noting that under certain
conditions the operator potentials ${\hat{\mathcal U}}_{\xi,w}(x)$
can be left alone in square brackets standing in front of the {\em
inter-mode} propagator $G_{mn'}$ in equation (\ref{ModEq-2}).
Indeed, the relative part of the terms quadratic in roughness
slope and the linear terms in those brackets is estimated by the
ratio
\begin{equation}
\frac{{\xi'}^2(x)}{w(x)\left| k\xi'(x)+\xi''(x)\right|}\sim
\frac{\sigma}{D(1+kr_c)} \ .
\label{quad/lin}
\end{equation}
It is evident that if the rms height of the boundary displacement
is small enough,
\begin{equation}
\sigma/D \ll 1 \ ,
\label{small_height}
\end{equation}
one can neglect the terms that are quadratic in $\xi'$ in
comparison with their adjacent linear counterparts. Besides, for
the sake of simplicity of the subsequent calculation we will
consider the boundary asperities to be also smooth,
\begin{equation}
\sigma/r_c \ll 1 \ ,
\label{small_slope}
\end{equation}
yet without requiring their adiabaticity in the sense of
Refs.~\cite{GlazLesKhmelShekh88,GlazJons91}.

The limitations (\ref{small_height}) and (\ref{small_slope}) are
common in solving the problems of classical wave scattering using
the perturbation theory\cite{BassFuks}. The restrictions are
motivated by the necessity of eliminating the well-known
``shadowing'' effect. In view of likeness of the mathematical
technique, this issue arises in the considered quantum problem as
well. However, under the conditions (\ref{small_height}) and
(\ref{small_slope}) there does not exist the wave shadowing in a
single-mode waveguide.

With all the above reasonings taken into account, we obtain the
equation for Green function $G_{11}(x,x')$ setting $n=n'=1$ in
Eq.~(\ref{ModEq-2}),
\begin{equation}
\left[ \frac{\partial^2}{\partial x^2} +
k_1^2 +i0 - V_h(x) - V_s(x)
\right] G_{11}(x,x')
- \sum_{m=2}^{\infty} {\hat U}_{1m}(x)
G_{m1}(x,x') = \delta(x-x') \ .
\label{SSB_1-Mod}
\end{equation}
Here we have introduced the notation $k^2_1$ for the
``unperturbed'' lengthwise energy of the extended mode $n=1$,
\begin{equation}
k_1^2 = k_F^2 - \left< \frac{\pi^2}{w^2(x)} \right> -
\left( 1 + \frac{\pi^2}{3} \right)
\left< \left[ \frac{w'(x)}{2w(x)} \right]^2 \right> \ .
\label{k_1}
\end{equation}
The quantities $V_h(x)$ and $V_s(x)$ in square brackets of
Eq.~(\ref{SSB_1-Mod}) stand for the {\em intra-mode} potentials
and have the form
\begin{subequations}
\label{VhVs}
\begin{eqnarray}
V_h(x) &=& \frac{\pi^2}{w^2(x)}  -
\left< \frac{\pi^2}{w^2(x)} \right> \ ,
\label{Vh} \\
V_s(x) &=& \left( 1 + \frac{\pi^2}{3} \right)
\left\{ \left[ \frac{w'(x)}{2w(x)} \right]^2 -
\left< \left[ \frac{w'(x)}{2w(x)} \right]^2 \right> \right\} \ .
\label{Vs}
\end{eqnarray}
\end{subequations}
Assuming these potentials to be taken as a perturbation, we design
them purposefully to have $\left<V_{h,s}(x)\right>=0$. As to the
{\em inter-mode} operator potential in Eq.~(\ref{SSB_1-Mod}),
\begin{equation}
{\hat U}_{1m}(x) = - B_{1m}\frac{2}{w(x)} {\hat{\mathcal
U}}_{w}(x) \ , \label{interSSB_1}
\end{equation}
it possesses this property by definition.

Although the mode Green functions in Eq.~(\ref{SSB_1-Mod}) depend
on a single space variable, the problem certainly cannot be
thought of as one-dimensional for coupling the function
$G_{11}(x,x')$ to all of the inter-mode propagators with mode
indices $m\neq 1$. Nevertheless, one can obtain a closed equation
for this function, which at weak scattering admits of the
perturbation analysis adequate to the diagrammatic method of
Refs.~\cite{berezinski73} and \cite{abrikryzh78}. In
Ref.~\cite{tar2000}, the evidence was given that in the case of
arbitrary 2D imperfect waveguide system governed by a set of
dynamic equations of the same functional structure as that of the
system (\ref{ModEq-2}), all of the inter-mode propagators
$G_{mn}(x,x')$ can be expressed, by means of a linear operator, in
terms of only one intra-mode Green function $G_{nn}(x,x')$.
Although for arbitrarily disordered systems this relation is of
little practical value for the complex dependence on scattering
potentials, under weak scattering conditions the reciprocal
operator expressions prove to be rather uncomplicated and readily
analyzable. Referring an interested reader to Ref.~\cite{tar2000}
for the exact procedure, we give here a simple recipe for
obtaining the approximate relation between the inter- and
intra-mode propagators, which is valid in the case of weak
electron scattering.

This recipe, already used in Ref.~\cite{MakTar}, consists in
solving the set of equations (\ref{ModEq-2}) iteratively with
respect to the inter-mode propagators entering equation
(\ref{SSB_1-Mod}). By letting $n'=1$ and re-designating the mode
variables one can reduce Eq.~(\ref{ModEq-2}) to a set of
inhomogeneous equations with respect to the functions
$G_{m1}(x,x')$ with $m>1$,
\begin{equation}
\hat{\mathcal G}_m^{-1} G_{m1}(x,x')- \sum_{k=2\atop(k\neq
m)}^{\infty}\hat U_{mk}(x)G_{k1}(x,x')= \hat U_{m1}(x)G_{11}(x,x')
\ . \label{Gm1->G11}
\end{equation}
It can be easily seen that all interesting propagators $G_{m1}$
can be expressed linearly through the single intra-mode propagator
$G_{11}$. In equation (\ref{Gm1->G11}), $\hat{\mathcal G}_m^{-1}$
is the differential operator from curly brackets of
Eq.~(\ref{ModEq-2}), where the term with ${{\xi'_c}^2}(x)$ should
be omitted, in accordance with the SSB model chosen, and
replacement of mode indices should be made $n\to m$. The
inter-mode potentials $\hat U_{mk}(x)$ in Eq.~(\ref{Gm1->G11})
have the form similar to the operator~(\ref{interSSB_1}),
\begin{equation}
{\hat U}_{mk}(x) = - B_{mk}\frac{2}{w(x)} {\hat{\mathcal
U}}_{w}(x) \ . \label{interSSB_m}
\end{equation}
%%%%%

It is essential that in the case of weak scattering all Green
functions ${\mathcal G}_m(x,x')$, with the aid of which the
equation (\ref{Gm1->G11}) is to be solved, belong to the class of
so-called {\em evanescent} functions. They are sufficient to be
taken in the unperturbed form (i.e. to zeroth order in boundary
roughness) which are found to be strongly localized,
\begin{eqnarray}
{\mathcal G}_m^{(0)}(x,x')= -\frac{1}{2|k_m|}
\exp\left(-|k_m||x-x'|\right) \ . \label{G_evan}
\end{eqnarray}
Here $|k_m|=\left[\left(\pi m/D\right)^2-k_F^2\right]^{1/2} > 0$.
Since at weak scattering all of the inter-mode potentials entering
Eq.~(\ref{Gm1->G11}) can be regarded as small (in functional
sense), the following approximate relation can be derived
iteratively,
\begin{equation}
G_{m1}(x,x')\approx\int_{-L/2}^{L/2}dx_1 {\mathcal
G}_m^{(0)}(x,x_1)\hat U_{m1}(x_1)G_{11}(x_1,x') \ .
\label{Gm1intG11}
\end{equation}
By substituting (\ref{Gm1intG11}) into (\ref{SSB_1-Mod}), we
arrive eventually at a closed equation for the intra-mode
propagator $G_{11}$,
\begin{equation}
\left[ \frac{\partial^2}{\partial x^2} + k_1^2 + i0 -V_h(x) - V_s(x)
\right] G_{11}(x,x')
-\int_{-L/2}^{L/2}dx_1\,\hat K(x,x_1)G_{11}(x_1,x') =
\delta(x-x') \ .
\label{G11_main}
\end{equation}

Equation (\ref{G11_main}) contains complete information on
scattering of single-extended mode electrons by the roughness of
the conductor boundaries. While local potentials $V_h(x)$ and
$V_s(x)$ are responsible for the intra-mode scattering, the
integral operator in (\ref{G11_main}) accounts also for the
inter-mode scattering. The kernel of this operator has the form
\begin{eqnarray}
{\hat K}(x,x') = - \sum_{m=2}^{\infty} B_{1m}^2 \left[ \left(
\frac{2}{w(x)} \right)^2 \right.  {\hat{\mathcal U}}_w(x){\mathcal
G}_m^{(0)}(x,x') {\hat{\mathcal U}}_w(x') - \left.\Big<\left(
\frac{2}{w(x)} \right)^2  {\hat{\mathcal U}}_w(x){\mathcal
G}_m^{(0)}(x,x') {\hat{\mathcal U}}_w(x') \Big> \right] \ ,
\label{K_inter}
\end{eqnarray}
being, in its turn, a differential operator. Similar to the
intra-mode potentials $V_{h,s}(x)$, the operator potential
(\ref{K_inter}) is constructed in such a way as to make
$\big<{\hat K}(x,x')\big>=0$. This certainly brings about even
greater complication of the exact form of the longitudinal energy
$k_1^2$ in comparison with that given in Eq.~(\ref{k_1}). But the
smoothness condition (\ref{small_slope}) makes it possible to omit
the terms containing the derivative $w'(x)$ in the ``unperturbed''
mode energy, thus replacing it with the simplified value
$k_1^2\approx k_F^2- \left<\pi^2/w^2(x)\right>$.

At last, in formula (\ref{Cond-mode}) for the conductance, in view
of the single-mode geometry of the conducting strip,
Eq.~(\ref{single-mode}), one should keep only the terms with the
diagonal mode propagator $G_{11}(x,x')$. With
Eqs.~(\ref{Phi_nm(x)}) and (\ref{small_slope}), the conductance
expression reduces to a relatively simple form,
\begin{eqnarray}
g(L) = - \frac{4}{L^2} \int_{-L/2}^{L/2}\!\!dx
\int_{-L/2}^{L/2}\!\!dx'\;
\frac{\partial G_{11}(x,x')}{\partial x}
\frac{\partial G_{11}^*(x,x')}{\partial x'} \ ,
\label{Cond_1-mode}
\end{eqnarray}
which will be the subject of further analysis in conjunction with
equation (\ref{G11_main}).

%--------------------------------------------
\section{Two-scale approach to the propagator calculation}
\label{Sec_two-scale}
%--------------------------------------------

Since we focus our attention on the single-mode quantum wires, the
unperturbed Green functions (\ref{G_evan}) of all higher modes
with $m\geq 2$ are real-valued and strongly localized in space.
The kernel of the operator potential in equation (\ref{G11_main})
is thus hermitian, which ensures hermicity of the total
Hamiltonian of the dynamic system.

Equation (\ref{G11_main}) can be solved asymptotically within WS
approximation. This approximation is most convenient to be stated
in terms of spatial lengths inherent in the problem. We will
arrange these lengths into two groups, viz. ``microscopic'' and
``macroscopic'' lengths. The electron Fermi wavelength, $2\pi
k_F^{-1}$, and the correlation length of boundary asperities,
$r_c$, are assigned to the first group. Among the macroscopic
lengths there will be scattering $\ell_{sc}$ of the extended mode
$n=1$ (to be determined below) and the length $L$ of an irregular
part of the conductor. In terms of these lengths, the criterion
for scattering to be classified as weak can be expressed through
the following inequalities,
\begin{equation}
k_F^{-1},r_c \ll \ell_{sc},L \ .
\label{weak-scatt}
\end{equation}
Note that correlation between the lengths pertaining to the same
group (either micro- or macroscopic) can be thought of as
arbitrary.

The retarded Green function entering the conductance expression
(\ref{Cond_1-mode}) is a solution to the \emph{boundary-value}
problem defined by equation (\ref{G11_main}) and the appropriate
boundary conditions. We assume the conducting strip to be open at
the ends $x=\pm L/2$ in the direction of the current. For an open
waveguide system, Sommerfeld's radiation conditions well-known in
classical wave theory \cite{BassFuks,Vladimirov67} are
appropriate. In the case of 1D equation (\ref{G11_main}) they can
be expressed in the Leontovich's form \cite{BassFuks,Klyatskin86}
\begin{equation}
\left.\left(\frac{\partial}{\partial x}\mp ik_1\right)
G_{11}(x,x')\right|_{x=\pm L/2}=0  \ ,
\label{1Dradcond}
\end{equation}
where the source is assumed to be placed inside the waveguide,
$x'\in [-L/2,L/2]$.

For lack of dynamic causality, the solution of Sturm-Liuville
problem (\ref{G11_main}), (\ref{1Dradcond}) is determined
functionally by the potentials in the bulk of the interval
$x\in(-L/2,L/2)$. This makes it rather difficult to obtain the
correlation functions by applying the well-elaborated methods of
statistical analysis which are applicable to the evolution-type
problems. One of the commonly used methods aimed at reducing
boundary-value problems to evolutionary ones is the invariant
imbedding method\cite{BellmanWing75,Klyatskin86}. Yet in this
study we apply a different approach which seems to be more
general. We will search for Green function of equation
(\ref{G11_main}) in the form
\begin{equation}
G_{11}(x,x')=\bm{\mathcal W}^{-1} \big[
\psi_+(x)\psi_-(x')\Theta(x-x') + \psi_+(x')\psi_-(x)\Theta(x'-x)
\big] \ , \label{Green-Cochy}
\end{equation}
where $\psi_{\pm}(x)$ are linearly independent solutions of the
homogeneous equation (\ref{G11_main}) supplemented with radiation
conditions analogous to (\ref{1Dradcond}) at only one of the strip
ends, $x=\pm L/2$, in accordance with the ``sign'' index of
$\psi_{\pm}$. The Wronskian of those functions is $\bm{\mathcal
W}$, $\Theta(x)$ is the Heaviside unit-step function.

It is advantageous for the further analysis to represent the
functions $\psi_{\pm}(x)$ as superpositions of modulated harmonic
waves propagating in opposite directions along the $x$-axis,
\begin{equation}
\psi_{\pm}(x) = \pi_{\pm}(x)\exp(\pm ik_1 x) -
i\gamma_{\pm}(x)\exp(\mp ik_1 x) \ .
\label{psi-pm}
\end{equation}
The WS approximation expressed in terms of inequalities
(\ref{weak-scatt}) suggests that the ``amplitudes'' $\pi_{\pm}(x)$
and $\gamma_{\pm}(x)$ in (\ref{psi-pm}) may be thought of as
varying slowly as compared to the ``fast'' exponentials $\exp(\pm
ik_1 x)$. This makes it possible to obtain the ``truncated''
equations for those amplitudes by averaging the exact
Schr\"{o}dinger equation over ``rapid'' phases, as it is done in
the theory of nonlinear oscillations\cite{BogolMitr74}.
Specifically, we multiply both sides of the homogeneous equation
(\ref{G11_main}) from the left by the exponent function $\exp(\mp
ik_1x)$ and then average all the terms over space interval $2l$,
which can be chosen to have the arbitrary length between the
micro- and macroscopic lengths of the problem,
\begin{equation}
k_F^{-1},r_c \ll l \ll \ell_{sc},L \ .
\label{interm_l}
\end{equation}
The final result is not expected to depend exactly on the choice
of the averaging interval.

By means of such a sub-averaging, we arrive at the following set
of first-order differential equations for the smooth amplitudes,
\begin{eqnarray}
&&\pm\pi'_{\pm}(x) + i\eta(x)\pi_{\pm}(x) +
\zeta_{\pm}^*(x)\gamma_{\pm}(x) =0\ ,
\nonumber\\[-.5\baselineskip]\label{Pi_Gamma}\\[-.5\baselineskip]\nonumber
&&\pm\gamma'_{\pm}(x) -
i\eta(x)\gamma_{\pm}(x) + \zeta_{\pm}(x)\pi_{\pm}(x)=0\ .
\end{eqnarray}
The radiation conditions for $\psi_{\pm}(x)$ are reformulated as
the following ``initial'' conditions for the functions $\pi_{\pm}$
and $\gamma_{\pm}$,
\begin{equation}
\pi_{\pm}(\pm L/2) = 1 \ , \qquad \qquad \gamma_{\pm}(\pm L/2) = 0 \ .
\label{In_cond}
\end{equation}
%

%--------------------------------------------
\section{Statistical properties of smoothed potentials}
%--------------------------------------------

At the last stage of developing the averaging procedure we specify
statistical properties of the functions $\eta(x)$ and
$\zeta_{\pm}(x)$ entering equations (\ref{Pi_Gamma}). These random
fields are defined as narrow packets of spatial harmonics of the
initial potentials from the lhs of Eq.~(\ref{G11_main}). The
contribution of intra-mode potential $V(x)=V_h(x)+V_s(x)$ to these
``smoothed'' potentials is represented by the expressions
\begin{equation}
\eta^{(V)}(x)=\frac{1}{2k_1}\int_{x-l}^{x+l}\frac{dt}{2l}V(t) \ ,
\hspace{1.5cm}
\zeta^{(V)}_{\pm}(x)=\frac{1}{2k_1}\int_{x-l}^{x+l}\frac{dt}{2l}
{\rm e}^{\pm 2ik_1t}V(t) \ . \label{EtaZeta}
\end{equation}
Unlike the local potential $V(x)$, the operator potential
$\hat{\mathcal K}$, which is specified by the kernel
(\ref{K_inter}), is, strictly speaking, non-local. Its part in the
random fields $\eta(x)$ and $\zeta_{\pm}(x)$ is described by more
complicated expressions than those given by Eq.~(\ref{EtaZeta}).
The corresponding potentials, the same as the kernel
(\ref{K_inter}) itself, are the differential operators. In
calculation of different correlation functions the random fields
$\eta(x)$ and $\zeta_{\pm}(x)$ are not of importance by
themselves, but only their statistical moments. Subject to WS
conditions, the latter can be calculated no matter what the
locality of the corresponding potential may be.

At weak scattering, all the potentials in Eq.~(\ref{G11_main}) may
be thought of as Gaussian distributed\cite{LifGredPas}.
Consequently, knowledge of binary correlators of those potentials
is sufficient to govern the statistics of all physical quantities.
It was shown in Ref.~\cite{MakTar} that under restrictions
(\ref{interm_l}) for the averaging interval $l$, only a pair of
correlators of the potentials $\eta(x)$ and $\zeta_{\pm}(x)$ are
not parametrically small, viz. $\left<\eta(x)\eta(x')\right>$ and
$\left<\zeta_{\pm}(x)\zeta_{\pm}^*(x')\right>$. Regardless of the
potential being local or not, calculation of these correlators
yields
\begin{subequations}
\label{Eta_Zeta}
\begin{eqnarray}
&&\left<\eta(x)\eta(x')\right>=
\frac{1}{L_f}F_l(x-x') \ ,
\label{EtaEta}\\
&&\left<\zeta_{\pm}(x)\zeta_{\pm}^*(x')\right>=
\frac{1}{L_b}F_l(x-x') \ .
\label{ZetaZeta}
\end{eqnarray}
\end{subequations}
The function
\begin{equation}
F_l(x)=\int_{-\infty}^{\infty}\frac{dq}{2\pi}{\text e}^{iqx}
\frac{\sin^2(ql)}{(ql)^2}=
\frac{1}{2l}\left(1-\frac{|x|}{2l}\right)\Theta(2l-|x|)
\label{F_l}
\end{equation}
in equations (\ref{Eta_Zeta}) is sharp in the scale of macroscopic
lengths, so hereinafter it can be regarded as the
$\delta$-function in the ``distributional'' sense,
$F_l(x)\to\delta(x)$.

The coefficients $L_{f,b}^{-1}$ in equations (\ref{Eta_Zeta})
represent the inverse scattering lengths, forward ($f$) and
backward ($b$), respectively. They are contributed by all the
potentials that result in both BH and BS scattering in equation
(\ref{G11_main}). In calculating these coefficients, it should be
borne in mind that under WS conditions one can disregard the
correlation between ``height'' potential (\ref{Vh}) and ``slope''
potentials (\ref{Vs}) and (\ref{K_inter})\cite{BassFuks}. This
permits us to perceive the potential (\ref{Vh}), on the one hand,
and potentials (\ref{Vs}) and (\ref{K_inter}), on the other, as
being associated with some different additive and mutually
non-interfering scattering mechanisms.

Scattering rates (both forward and backward) associated with
``slope'' potentials similar to (\ref{Vs}) and (\ref{K_inter})
were previously studied in Ref.~\cite{MakTar} for CCB model of a
roughly bounded strip. Although the CCB waveguide is somewhat
different from the SSB waveguide addressed here, the difference
appears to arise only in numerical factors at the potentials
considered in paper \cite{MakTar} and potentials (\ref{Vs}) and
(\ref{K_inter}) of this work. Therefore, referring the reader to
Ref.~\cite{MakTar} for technical details, we present here the
final expressions for the inverse scattering lengths associated
with BS scattering, which are valid for the SSB waveguide model,
\begin{subequations}
\label{slope_scatt}
\begin{eqnarray}
\frac{1}{L_f^{(s)}} =
\frac{1}{2k_1^2}\left(\frac{\sigma}{D}\right)^4
\int_{-\infty}^\infty \frac{dq}{2\pi}\, q^4 \,\widetilde{\mathcal
W}^2(q) \Bigg\{&& \left(1+\frac{\pi^2}{3} \right) \nonumber \\
&&+ 2\sum\limits_{m=2}^\infty B_{1m}^2 \left[ (2k_1+q)^2
\widetilde{\mathcal G}_{m}^{(0)}(k_1+q) + (2k_1-q)^2
\widetilde{\mathcal G}_{m}^{(0)}(k_1-q) \right]\Bigg\}^2 \ ,\qquad
\label{L_f(s)}
\end{eqnarray}
\begin{eqnarray}
\frac{1}{L_b^{(s)}} =
\frac{1}{2k_1^2}\left(\frac{\sigma}{D}\right)^4
\int_{-\infty}^\infty \frac{dq}{2\pi}\, (q^2-k_1^2)^2 \, %&&
\widetilde{\mathcal W}(q-k_1)\widetilde{\mathcal W}(q+k_1)
\left[\left(1+\frac{\pi^2}{3}\right) +4\sum\limits_{m=2}^\infty
B_{1m}^2(q^2-k_1^2) \widetilde{\mathcal G}_m^{(0)}(q)\right]^2 \
.\qquad \label{L_b(s)}
\end{eqnarray}
\end{subequations}
The functions $\widetilde{\mathcal W}(q)$ and $\widetilde{\mathcal
G}_m^{(0)}(q)$ in Eqs.~(\ref{slope_scatt}) are the Fourier
transforms of the correlation function ${\mathcal W}(x)$ from
(\ref{xi-corrCCB}) and the evanescent Green function
(\ref{G_evan}), respectively.

As regards the BH scattering due to the potential (\ref{Vh}), the
corresponding scattering rate is worth considering here in more
detail. At first glance, it seems natural to find the
corresponding frequency by expanding the potential (\ref{Vh}),
with condition (\ref{small_height}), in small fluctuations of the
conductor width,
\begin{equation}
V_h(x) \approx -\frac{2\pi^2}{D^3}\Delta w(x) \ .
\label{Vh_if}
\end{equation}
However, in trying to improve the obtained scattering frequency by
retaining the terms of higher order in $\Delta w(x)$ it turns out
that the corresponding series converges non-uniformly, so that for
its convergence it is necessary to hold a great number of terms.
The similar problem was encountered earlier in
Ref.~\cite{Kawabata93}, where the artificial ``cutting'' parameter
was introduced for the corresponding series to become ultimately
convergent.

This difficulty can be overcome if the expansion of the BH
potential in series of small displacement of the confining
surfaces is discarded, but the exact expression (\ref{Vh}) is used
instead. This can be easily done for the gaussian roughness model,
the appropriate scheme being presented in the Appendix. With this
technique, one can obtain for inverse scattering lengths
pertaining to the potential (\ref{Vh}) the expressions below,
which are valid in the case of small boundary asperities,
Eq.~(\ref{small_height}),
\begin{subequations}
\label{L_fhL_bh}
\begin{eqnarray}
\frac{1}{L_f^{(h)}} &=& \frac{4\pi^4\sigma^2}{k_1^2D^6}
\widetilde{\mathcal W}(0) \ , \label{L_fh} \\ \frac{1}{L_b^{(h)}}
&=& \frac{4\pi^4\sigma^2}{k_1^2D^6} \widetilde{\mathcal W}(2k_1) \
. \label{L_bh}
\end{eqnarray}
\end{subequations}
It is noteworthy that extinction lengths (\ref{L_fhL_bh}), being
obtained by rigorous calculation, coincide exactly in form with
those obtained by means of the lowest-order expansion
(\ref{Vh_if}) of the Hamiltonian (\ref{Vh}) in powers of $\Delta
w(x)$.

%--------------------------------------------
\section{Conductance and resistivity moments}
\label{cond_mom}
%--------------------------------------------

To perform the statistical averaging based on the effective
zero-scale correlation of random potentials (\ref{Eta_Zeta}) it is
necessary to express the conductance (\ref{Cond_1-mode}) in terms
of smooth amplitudes $\pi_{\pm}$ and $\gamma_{\pm}$. Under WS
conditions (\ref{weak-scatt}), when substituting the Green
function in the form (\ref{Green-Cochy}) into the expression
(\ref{Cond_1-mode}), it is sufficient to differentiate over
coordinate variables only fast exponentials in wave functions
(\ref{psi-pm}). As a result, the conductance takes on the
intermediate form
\begin{eqnarray}
g(L) = \frac{4k_1^2}{L^2|{\cal W}|^2} \int_{-L/2}^{L/2} dx
\Bigg\{&& \left[|\pi_+(x)|^2 - |\gamma_+(x)|^2\right]
\int_{-L/2}^{x} dx' \left[|\pi_-(x')|^2 - |\gamma_-(x')|^2\right]
\nonumber \\ && + \left[|\pi_-(x)|^2 - |\gamma_-(x)|^2\right]
\int_{x}^{L/2} dx' \left[|\pi_+(x')|^2 - |\gamma_+(x')|^2\right]
\Bigg\} \ . \label{g-pi-gamma}
\end{eqnarray}
Within the same accuracy, the Wronskian ${\cal W}$ in
(\ref{g-pi-gamma}) equals
\begin{equation}
{\cal W}\approx 2ik_1\left[ \pi_+(x)\pi_-(x) +
\gamma_+(x)\gamma_-(x)\right] = 2ik_1 \pi_\pm(\mp L/2) \ ,
\label{W-2scale}
\end{equation}
where the last equality is a consequence of ``boundary''
conditions (\ref{In_cond}), which are valid within WS
approximation as well.

Formula (\ref{g-pi-gamma}) can be simplified if the symmetry
properties of equation (\ref{G11_main}) and hermicity of the
corresponding 1D Hamiltonian are taken into account. Since the
problem (\ref{Pi_Gamma}), (\ref{In_cond}) is of evolutionary type,
its solution can be represented in terms of an $x$-ordered matrix
exponential,
\begin{equation}
{\bf I}_\pm(x) = \hat T_x \exp\bigg[\pm \int_{x}^{\pm L/2} dx'
{\bf b}(x') \bigg] \ . \label{I=solv}
\end{equation}
Here the matrices of smooth amplitudes, ${\bf I}_\pm(x)$, have the
form
\begin{equation}
{\bf I}_+(x) = \left(
\begin{array}{cc}
\pi_+(x) & \gamma_+(x) \\ \gamma_+^*(x) & \pi_+^*(x)
\end{array}
\right) \ , \qquad \qquad {\bf I}_-(x) = \left(
\begin{array}{cc}
\pi_-(x) & \gamma_-^*(x) \\ \gamma_-(x) & \pi_-^*(x)
\end{array}\right) \ ,
\label{I_pm=def}
\end{equation}
${\bf b}(x)$ is the random field matrix,
\begin{equation}
{\bf b}(x) = \left(
\begin{array}{cc}
i\eta(x) & \zeta_{+}(x) \cr  \zeta_{-}(x) & -i\eta(x)
\end{array}\right) \ ,
\label{b=def}
\end{equation}
whose off-diagonal elements are interconnected by the equality
$\zeta_{-}(x)=\zeta_{+}^*(x)$. The operator $\hat T_x$ in
Eq.~(\ref{I=solv}) arranges the multipliers in each of the terms
of the exponential series in order of decreasing their coordinate
arguments from left to right.

Taking advantage of the operator identity $\ln\det{\bf A} \equiv
{\rm Sp}\ln{\bf A}$, matrices (\ref{I_pm=def}) can be shown to be
unimodular,
\begin{equation}
\det {\bf I}_\pm(x) = |\pi_\pm(x)|^2 - |\gamma_\pm(x)|^2 = 1 \ .
\label{Unimod}
\end{equation}
This relation, along with equality (\ref{W-2scale}), results in
the following form of the conductance of a single-mode wire,
\begin{equation}
g(L)=\left| \pi_\pm^{-1}(\mp L/2) \right|^2 \ . \label{cond-fin}
\end{equation}
Holding to Landauer's concept, the quantity $\pi_\pm^{-1}(\mp
L/2)$ is to be interpreted as the transmission coefficient of a
single-mode quantum waveguide of length $L$. This interpretation
is supported by the following argumentation. From the structure of
wave functions (\ref{psi-pm}) it follows that the ratio
$\Gamma_+(x)=\gamma_+(x)/\pi_+(x)$ is defined as the reflection
coefficient for the harmonics $k_1$ incident onto the interval
$(x,L/2)$ with a unit amplitude from the left-hand side.
Correspondingly, $\Gamma_-(x)=\gamma_-(x)/\pi_-(x)$ represents the
reflection coefficient of the harmonics $-k_1$ incident onto the
interval $(-L/2,x)$ from the right-hand side. Subject to this
definition, equation (\ref{Unimod}) can be re-written in the form
of the conservation law in a non-dissipative medium,
\begin{equation}
|\Gamma_\pm(x)|^2 + |\pi_\pm^{-1}(x)|^2 = 1 \ ,
\label{flux-conserv}
\end{equation}
whereupon the interpretation of quantity (\ref{cond-fin}) as a
square modulus of the transmission coefficient of the disordered
interval $(-L/2,L/2)$ seems to be apparent.

With the conservation law (\ref{flux-conserv}), it is convenient
to perform subsequent calculations of the statistical moments of
the conductance using the equation for $\Gamma_\pm(x)$ rather than
that for the transmission coefficient $\pi_\pm^{-1}(x)$. From
Eqs.~(\ref{Pi_Gamma}), the reflection coefficient can be found to
obey the Riccati-type closed equation subject to the zero initial
condition,
\begin{eqnarray}
\pm && \frac{d\Gamma_{\pm}(x)}{dx} = 2i\eta(x)\Gamma_{\pm}(x) +
\zeta_{\pm}^*(x)\Gamma_{\pm}^2(x) - \zeta_{\pm}(x) \ ,
\label{Gam-eq} \\[6pt] && \ \Gamma_{\pm}(\pm L/2) = 0 \ . \nonumber
\end{eqnarray}
The forward-scattering random field $\eta(x)$ may be eliminated
from Eq.~(\ref{Gam-eq}) by concurrent phase transformation of the
function $\Gamma_\pm(x)$ and the backscattering field
$\zeta_{\pm}(x)$,
\begin{subequations}
\label{phase-transform}
\begin{eqnarray}
\Gamma_\pm(x) &=& \Gamma_\pm^{(new)}(x) \exp\left[ \pm 2i
\int_{\pm L/2}^{x} dx' \eta(x') \right] \ , \label{Gamma-def} \\
\zeta_{\pm}(x)&=&\zeta_{\pm}^{(new)}(x)\exp\left[\pm 2i \int_{\pm
L/2}^{x} dx' \eta(x') \right] \ . \label{zeta-transform}
\end{eqnarray}
\end{subequations}
This transformation keeps the conductance (\ref{cond-fin}) and
correlation relation (\ref{ZetaZeta}) unaffected, so one may put
function $\eta(x)$ in Eq.~(\ref{Gam-eq}) equal to zero. As a
consequence, the outcome for an arbitrary moment of the
conductance is to be specified exclusively by the {\em
backscattering} of the electrons, i.e. it will depend on the
scattering length which inverse value is the sum of inverse
lengths (\ref{L_b(s)}) and (\ref{L_bh}).

To proceed further, consider the $n$th moment of the local
reflection coefficient squared modulus,
\begin{equation}
R_n^{\pm}(x)=\langle |\Gamma_{\pm}(x)|^{2n}\rangle \ .
\label{R_n-def}
\end{equation}
Since the stochastic problem (\ref{Gam-eq}) is of the evolutionary
type, it can be reduced, via Furutsu-Novikov
formalism\cite{Klyatskin86}, to the differential-difference
equation for moments (\ref{R_n-def}) (see also
Ref.~\cite{AshKohlerPap91}),
\begin{equation}
\pm\frac{dR_n^{\pm}(x)}{dx} = -\frac{n^2}{L_b} \left[
R_{n+1}^{\pm}(x)-2R_n^{\pm}(x)+R_{n-1}^{\pm}(x)\right],
\qquad\qquad n=0,1,2,\ldots \ . \label{R_n-eq}
\end{equation}
Here $L_b^{-1}=L_b^{(s){-1}}+L_b^{(h){-1}}$. The initial condition
on the coordinate $x$ to the solution of Eq.~(\ref{R_n-eq}) is
\begin{equation}
R_n^{\pm}(\pm L/2) = \delta_{n0}. \label{R_n-cond}
\end{equation}
As for the dependence of $R_n^{\pm}(x)$ on the discrete variable
$n$, it follows from the definition (\ref{R_n-def}) that
$R_0^{\pm}(x)=1$ and $R_n^\pm(x) \to 0$ as $n\to\infty$.

The solution of Eq.~(\ref{R_n-eq}) that matches all the above
mentioned conditions can be expressed through the probability
function $P_L^\pm(u|x)$ and represented, upon due parametrization,
in the form
\begin{equation}
R_n^{\pm}(x) = \int_1^\infty du P_L^\pm(u|x)
\left(\frac{u-1}{u+1}\right)^n. \label{R_n-int}
\end{equation}
Correspondingly, the statistical moments of the conductance
(\ref{cond-fin}) are represented by the following integral,
\begin{equation}
\big< g^n(L)\big> = \big< \left(1-|\Gamma_{\pm}(\mp
L/2)|^2\right)^n \big> = \int_1^\infty du P_L^\pm(u|\mp L/2)
\left({2 \over u+1}\right)^n. \label{g^n-int}
\end{equation}

To obtain the probability density $P_L^\pm(u|x)$ one should
substitute $R_n^{\pm}(x)$ in the form (\ref{R_n-int}) into
equation (\ref{R_n-eq}), thus obtaining the Fokker-Plank equation,
\begin{equation}
\pm L_b\frac{\partial P_L^\pm(u|x)}{\partial x} =
-\frac{\partial}{\partial u}(u^2-1) \frac{\partial
P_L^\pm(u|x)}{\partial u} \ , \label{Fokk-Plank}
\end{equation}
which is supplemented, according to Eq.~(\ref{R_n-cond}), by the
initial condition on the coordinate $x$,
\begin{equation}
P_L^\pm(u|\pm L/2) = \delta(u-1-0) \ . \label{P_L-ic}
\end{equation}
Besides, normalization of the function $P_L^\pm(u|x)$ to unity is
ensured by $R_0^{\pm}(x)=1$. This implies that the distribution
function is integrable over the variable $u$, in particular, at
$u\to 1$ and $u\to\infty$.

The solution to Eq.~(\ref{Fokk-Plank}), which meets the above
mentioned requirements, can be found by using the Mehler-Fock
transformation and has the conventional form \cite{LifGredPas}
\begin{eqnarray}
P_L^\pm(\cosh\alpha|x) =  \frac{1}{\sqrt{8\pi}} \left( \frac{L \mp
2x}{2L_b} \right)^{-3/2} \exp \left(-\frac{L \mp 2x}{8L_b} \right)
\int_{\alpha}^{\infty} \frac{v\,dv}{(\cosh v - \cosh\alpha)^{1/2}}
\exp \left[ -\frac{v^2}{4} \left(\frac{L \mp 2x}{2L_b}
\right)^{-1} \right] \ ,\qquad\quad \label{P_L}
\end{eqnarray}
where the change of a variable has been made $u = \cosh\alpha$,
$\alpha \geq 0$. With this expression, equality (\ref{g^n-int})
yields the relatively simple (as well as suitable to analyze)
formula for the $n$th moment of the dimensionless conductance,
\begin{eqnarray}
\big< g^n(L)\big> &= & \frac{4}{\sqrt\pi}
\left(\frac{L_b}{L}\right)^{3/2}
\exp{\left(-\frac{L}{4L_b}\right)}
\int_0^\infty\frac{zdz}{\cosh^{2n-1}z}
\exp\left(-z^2\frac{L_b}{L}\right) \int_0^z dy\, \cosh^{2(n-1)} y
\, , \label{g-moments}
\\[10pt]
 n &=& 0,\pm 1,\pm2,\ldots  \ . \nonumber
\end{eqnarray}

The result (\ref{g-moments}) completely determines main averaged
transport characteristics of a single-mode conducting strip. In
particular, although the conductance itself is not a self-averaged
quantity, one can, in principle, calculate its self-averaged
logarithm using the whole set of statistical moments
(\ref{g-moments}). However, it is much easier to obtaine this
quantity directly from equations (\ref{Pi_Gamma}), omitting
cumbersome manipulations with a logarithmic series of the terms
(\ref{g-moments}). Specifically, by differentiating the quantity
$|\pi_{\pm}^{-1}(x)|^2$ over $x$ we arrive at the equation
\begin{equation}
\pm\frac{d}{dx} \ln|\pi_{\pm}^{-1}(x)|^2 =
\zeta_{\pm}^*(x)\Gamma_{\pm}(x) + \zeta_{\pm}(x) \Gamma_{\pm}^*(x)
\ . \label{eq_lncond}
\end{equation}
By integrating Eq.~(\ref{eq_lncond}) over the interval
$(-L/2,L/2)$, with (\ref{In_cond}) taken into account, we
immediately obtain the logarithm of the conductance in the
left-hand side. Before averaging the terms in the rhs of
Eq.~(\ref{eq_lncond}), we point out that from equation
(\ref{Gam-eq}) it follows that function $\Gamma_{\pm}(x)$, being
considered as a functional of random fields $\zeta_{\pm}$ and
$\zeta_{\pm}^*$, depends on the value of these fields exactly
within the interval $(x,\pm L/2)$, according to the sign
index~$(\pm)$. Since at weak scattering all the above mentioned
random fields can be regarded as Gaussian distributed functional
variables, we can apply the Furutsu-Novikov formalism for its
averaging\cite{Klyatskin86}. The average of the first term in the
rhs of Eq.~(\ref{eq_lncond}) can, therefore, be presented in the
form
\begin{equation}
\left< \zeta_{\pm}^*(x) \Gamma_{\pm}(x)  \right> = \pm\int_x^{\pm
L/2} dx' \left< \zeta_{\pm}^*(x)\zeta_{\pm}(x') \right>
\Big<\frac{\partial\Gamma_{\pm}(x)}{\partial\zeta_{\pm}(x')}\Big>=
\frac{1}{2L_b}\left.
\Big<\frac{\partial\Gamma_{\pm}(x)}{\partial\zeta_{\pm}(x')}\Big>
\right|_{x'\to x\pm 0} \ , \label{FurNov}
\end{equation}
where we have used the effective $\delta$-correlation of the
fields $\zeta_{\pm}(x)$ and $\zeta_{\pm}^*(x)$. The variational
derivative in (\ref{FurNov}) can be readily calculated with the
use of equation (\ref{Gam-eq}), and thus it turns out to be equal
exactly to unity.  Finally, we arrive at the well-known result for
1D disordered systems,
\begin{equation}
\left<\ln g(L)\right>=-L/L_b \ , \label{ln_g-av}
\end{equation}
signaling the exponential fall of the conductance with a growing
length $L$ at the so-called representative (non-resonant)
realizations of the random potential\cite{LifGredPas}.

%--------------------------------------------
\section{Discussion of the results}
\label{discuss}
%--------------------------------------------

With general formula (\ref{g-moments}), write down the expressions
for average resistance $\langle g^{-1}(L) \rangle$ and average
conductance $\langle g(L) \rangle$. At $n=-1$ the integrals in
Eq.~(\ref{g-moments}) can be calculated exactly, so the average
resistance is equal to
\begin{equation}
\langle g^{-1}(L) \rangle = {1 \over 2}\left[1+\exp\left({2L\over
L_b}\right)\right] \ . \label{sopr}
\end{equation}
At $n=1$ the integration can be performed asymptotically in the
parameter $L/L_b$, giving rise to the following expression:
\begin{equation}
\langle g(L) \rangle \approx \left\{
\begin{array}{ll}
1-L/L_b &\mbox{ if } \qquad L/L_b\ll 1   \\[6pt]
2^{-1}\pi^{5/2}\left(L/L_b\right)^{-3/2} \exp\left(-L/ 4L_b\right)
\qquad &\mbox{ if } \qquad L/L_b\gg 1 \ .
\end{array}
\right. \label{cond_assymp}
\end{equation}

The results (\ref{ln_g-av})--(\ref{cond_assymp}) are completely in
line with the concepts of the localization theory for
one-dimensional disordered systems. Obvious indications of the
ballistic electron transport in short wires can be easily seen,
$g(L)\approx 1$ at $L\ll L_b$. Also, no signs of diffusive motion
of the electrons in long wires are present in any result. And
conversely, in long wires, $L\gg L_b$, the resistance (\ref{sopr})
displays an exponential increase with a growing strip length, and
the asymptotic (\ref{cond_assymp}) shows an exponential decrease
of the average conductance as the length $L$ exceeds the value of
$4L_b$. That behaviour is characteristic for conduction electrons
undergoing Anderson localization. The inverse of the quadruple
backscattering extinction length is equal to the Lyapunov exponent
for the electron wave function in the case of dimension
unity\cite{LifGredPas}, so that the quantity $\ell_{\rm loc}=4L_b$
is conventionally called the (one-dimensional) localization
length.

Both of the formulas (\ref{g-moments}) and (\ref{ln_g-av}), and,
hence, (\ref{sopr}) and (\ref{cond_assymp}), are universal in that
they are applicable for any one-dimensional degenerate system
subject to weak \emph{static} disorder. Only the particular
dependence of scattering lengths $L_{f,b}$ and, consequently, the
localization length $\ell_{\rm loc}$ is determined by the physical
nature of disorder. In this paper, we have found that if the
boundary roughness prove to be just the main cause of the disorder
in a 2D single-mode conducting strip, the interpretation of
scattering mechanism may be substantially different, depending on
the interrelation between by-slope scattering lengths
(\ref{slope_scatt}) and by-height lengths (\ref{L_fhL_bh}). To
make the correct comparison of the lengths the roughness
statistics needs to be specified. We will make the comparison for
two characteristic models consistent with the demand for
analyticity of the random functions $\xi_{1,2}(x)$ descriptive of
boundaries of the conductor. Specifically, we examine the case of
asperities subject to gaussian (exponential) correlation
statistics, ${\cal W}(x)= \exp\left(-x^2/2r_c^2 \right)$, an those
described by the Lorentz (power-type) correlation function ${\cal
W}(x)=\left[1+(x/r_c)^2\right]^{-1}$. For the gaussian roughness,
from (\ref{L_b(s)}) and  (\ref{L_bh}) the estimates can be
obtained
\begin{subequations}
\label{LbGauss}
\begin{equation}
\left.\frac{1}{L_b^{(h)}}\right|_{\rm{Gauss}}
\sim\left(\frac{\sigma}{D}\right)^2 \frac{r_c}{D^2}\left\{
\begin{array}{lcl}
1 &\qquad\rm{if}\quad& r_c/D\ll 1 \ (\rm{i.e.}\ k_1r_c\ll 1) \\[6pt]
\exp\left(-2k_1^2r_c^2\right) & \qquad\rm{if}\quad& r_c/D\gg 1 \
(\rm{i.e.}\ k_1r_c\gg 1) \ ,
\end{array}  \right.
\label{LbhGauss}
\end{equation}
\begin{equation}
\left.\frac{1}{L_b^{(s)}}\right|_{\rm{Gauss}}
\sim\left(\frac{\sigma}{r_c}\right)^4 \frac{r_c}{D^2}\left\{
\begin{array}{lcl}
1 &\qquad\rm{if}\quad& r_c/D\ll 1 \ (\rm{i.e.}\ k_1r_c\ll 1) \\[6pt]
(k_1r_c)^4\exp\left(-k_1^2r_c^2\right) & \qquad\rm{if}\quad&
r_c/D\gg 1 \ (\rm{i.e.}\ k_1r_c\gg 1) \ .
\end{array}  \right. \qquad\quad
\label{LbsGauss}
\end{equation}
\end{subequations}
If the hight correlation is lorentzian, the estimates change to
the following,
\begin{subequations}
\label{LbLorentz}
\begin{equation}
\left.\frac{1}{L_b^{(h)}}\right|_{\rm{Lorentz}}
\sim\left(\frac{\sigma}{D}\right)^2 \frac{r_c}{D^2}\left\{
\begin{array}{lcl}
1 &\qquad\rm{if}\quad & r_c/D\ll 1 \ (\rm{i.e.}\ k_1r_c\ll 1) \\[6pt]
\exp\left(-2k_1r_c\right) & \qquad\rm{if}\quad& r_c/D\gg 1 \
(\rm{i.e.}\ k_1r_c\gg 1) \ ,
\end{array}  \right.
\label{LbhLorentz}
\end{equation}
\begin{equation}
\left.\frac{1}{L_b^{(s)}}\right|_{\rm{Lorentz}}
\sim\left(\frac{\sigma}{r_c}\right)^4 \frac{r_c}{D^2}\left\{
\begin{array}{lcl}
1 &\qquad\rm{if}\quad& r_c/D\ll 1 \ (\rm{i.e.}\ k_1r_c\ll 1) \\[6pt]
(k_1r_c)^5\exp\left(-2k_1r_c\right) & \qquad\rm{if}\quad& r_c/D\gg
1 \ (\rm{i.e.}\ k_1r_c\gg 1) \ .
\end{array}  \right. \qquad\quad
\label{LbsLorentz}
\end{equation}
\end{subequations}

Based on estimates (\ref{LbGauss}) and (\ref{LbLorentz}), the
relative intensity of BS and BH scattering can be estimated as
follows,
\begin{subequations}
\label{Lh/Lb}
\begin{equation}
\left.\frac{L_b^{(h)}}{L_b^{(s)}}\right|_{\rm{Gauss}}
\sim\left(\frac{\sigma}{D}\right)^2
\left(\frac{D}{r_c}\right)^4\left\{
\begin{array}{lcl}
1 &\qquad\rm{if}\quad& r_c/D\ll 1 \ (\rm{i.e.}\ k_1r_c\ll 1) \\[6pt]
(r_c/D)^4\exp\left(k_1^2r_c^2\right) & \qquad\rm{if}\quad&
r_c/D\gg 1 \ (\rm{i.e.}\ k_1r_c\gg 1) \ ,
\end{array}  \right. \qquad\quad
\label{Lh/LbGauss}
\end{equation}
\begin{equation}
\quad\left.\frac{L_b^{(h)}}{L_b^{(s)}}\right|_{\rm{Lorentz}}
\sim\left(\frac{\sigma}{D}\right)^2
\left(\frac{D}{r_c}\right)^4\left\{
\begin{array}{lcl}
1 &\qquad\rm{if}\quad& r_c/D\ll 1 \ (\rm{i.e.}\ k_1r_c\ll 1) \\[6pt]
(r_c/D)^5 & \qquad\rm{if}\quad& r_c/D\gg 1 \ (\rm{i.e.}\ k_1r_c\gg
1) \ .
\end{array}  \right.
\label{Lh/LbLorentz}
\end{equation}
\end{subequations}
From these estimations it can be seen that in all of the limiting
cases considered here the characteristic ratio
$L_b^{(h)}/L_b^{(s)}$ is determined by the product of small
Rayleigh parameter $\left(\sigma/D\right)^2$ and some scale
parameter, though individual for different roughness correlation,
which depends on the relation between the roughness correlation
length and de-Broglie wave length of the electrons. In the case of
small-scale asperities, when $k_Fr_c\ll 1$ (or, which is the same,
$r_c/D\ll 1$), regardless of the correlation model the relative
intensity of BH and BS scattering is characterized by the
parameter $\left(\sigma/
D\right)^2(D/r_c)^4=(\sigma/r_c)^4/(\sigma/D)^2$ which varies over
a rather wide range. This is because of BH and BS scattering being
associated with independent physical origins. Whereas scattering
from the potential $V_h(x)$ is governed by the height of the
boundary roughness and thus is estimated mostly in terms of the
parameter $\sigma/D$, Eqs.~(\ref{LbhGauss}) and
(\ref{LbhLorentz}), the BS scattering is mostly determined by the
slope of the asperities, i.e. by gradients $\xi_{1,2}'(x)$, and
is, consequently, governed by the parameter $\sigma/r_c$,
Eqs.~(\ref{LbsGauss}) and (\ref{LbsLorentz}).

It would be tiresome to discuss here in detail the interrelation
between BH and BS scattering mechanisms assuming the asperities to
be large-scale, when the parameter $k_Fr_c\gg 1$ (i.e. $r_c/D\gg
1$), since in this case the the result depends largely on the
correlation model. We leave this particular analysis to an
interested reader.

It is noteworthy that in the event when the ``slope'' mechanism
dominates the ``height'' mechanism, the ``surface'' scattering
rate is proportional to the forth power of the rms height $\sigma$
rather than to $\sigma^2$, which is customary in the diffraction
theory\cite{BassFuks}. This fact must be taken into account when
analyzing experiments aimed at reproducing the surface shape using
the data on quantum, as well as classical, wave scattering in
rough-bounded waveguide systems.

Apart from the identification of dominant scattering mechanism,
specifying scattering lengths (\ref{LbGauss}) and
(\ref{LbLorentz}) for different statistical models of boundary
roughness also allows for the criteria of validity of the obtained
results in terms of essential physical parameters of the
disordered systems. Along with the presumption of smallness and
smoothness of boundary asperities (see Eqs.~(\ref{small_height})
and (\ref{small_slope})) the criteria are dictated by
(\ref{weak-scatt}) of weak scattering. In specifying those
criteria, the smallest of the extinction lengths $L_f^{(h,s)}$
should be taken to substitute the length $\ell_{sc}$ in
(\ref{weak-scatt}) since the inequality always holds true
$L_f\lesssim L_b$.

In conclusion, we make some remarks concerning the methodological
side of the problem of wave scattering from rough waveguide
surfaces. As far as we know, until the present time there has not
been made any reasonable distinction between BH and BS scattering
in such systems. Only the existence in general of different
competing mechanisms responsible for wave scattering from rough
surfaces was indicated in Ref.~\cite{DietzeDarling96} on the basis
of the experimental results. Accordingly, the relative function of
these scattering mechanisms in dynamic processes in waveguide-like
systems was not properly analyzed. Meanwhile, in the course of
this work we have made certain that application of linearized
(impedance-type) boundary conditions to a single-mode waveguide is
equivalent to retaining in Eq.~(\ref{G11_main}) the approximate
potential (\ref{Vh_if}) instead of its exact value (\ref{Vh}), and
also disregarding all ``slope'' potentials. However, omitting the
latter potentials implies the neglect of BS scattering mechanism,
that is proven to be not always justifiable. Alternative
small-slope approximation of
Refs.~\cite{Voronovich94,Voronovich96}, being guided by a solely
slope parameter of the rough surface, does not allow one to
separate BH and BS scattering mechanisms as well.

In this work, the method has been suggested within the framework
of which both of the above-mentioned scattering mechanisms appear
quite naturally, being associated with different terms of the
Hamiltonian. We have demonstrated that at least in a single-mode
waveguide the scattering caused even by mildly sloping boundary
asperities can be attributed to either ``height'' or ``slope''
scattering mechanism, depending on the statistical properties of
the roughness. The competition between these mechanisms is
governed by physically different parameters. It is noteworthy that
taking into account the BS mechanism is particularly essential if
boundary asperities are classified as being large-scale.

\begin{acknowledgements}
N.M.M. acknowledges support from CONACYT.
\end{acknowledgements}

%%%%%%%%%%%%%%%%%%%%%%%%%%%%%%%%%%%%%%%%%%%
\appendix
%%%%%%%%%%%%%%%%%%%%%%%%%%%%%%%%%%%%%%%%%%%

%------------------------------------------
\section*{Appendix: Correlation characteristics of by-height scattering}
\label{ByHeight}
%------------------------------------------

The effective sub-averaged potentials $\eta_h(x)$ and
$\zeta_{h\pm}(x)$ corresponding to the original potential
(\ref{Vh}) have the form
\begin{subequations}
\label{Eta_hZeta_h}
\begin{eqnarray}
&& \eta_h(x)=\frac{\pi^2}{2k_1}\int_{x-l}^{x+l}\frac{dt}{2l} \;
\left[ \frac{1}{w^2(t)}  - \left< \frac{1}{w^2(t)} \right> \right]
\ , \label{Eta_h1} \\ &&
\zeta_{h\pm}(x)=\frac{\pi^2}{2k_1}\int_{x-l}^{x+l}\frac{dt}{2l} \;
{\rm e}^{\mp 2ik_1t}\left[ \frac{1}{w^2(t)}  - \left<
\frac{1}{w^2(t)} \right> \right] \ . \label{Zeta_h1}
\end{eqnarray}
\end{subequations}
To calculate the averages in Eqs.~(\ref{Eta_hZeta_h}), it is
convenient to make use of the method below,
\begin{eqnarray}
\left< \frac{1}{w^2(t)} \right> &=& -\frac{\partial}{\partial D}
\left< \frac{1}{w(t)} \right> \ , \label{Av1/w} \\ \left<
\frac{1}{w(t)} \right> = \left< \frac{1}{D+\Delta\xi(t)} \right>
&=& -i\int_0^{\infty} dq \; \Big< \exp\left\{
iq[D+\Delta\xi(t)+i0] \right\} \Big> \ . \label{Av1/w2}
\end{eqnarray}
In the case of boundary roughness obeying gaussian statistics, the
integral in (\ref{Av1/w2}) can be averaged without difficulty.
Since for the SSB waveguide the equality holds true
\begin{equation}
\langle \Delta\xi(x)\Delta\xi(x') \rangle = 4\sigma^2 {\cal
W}(x-x') \ , \label{SSBcorr}
\end{equation}
the correlator (\ref{Av1/w2}) equals
\begin{equation}
\left< \frac{1}{w(t)} \right> =
-\frac{i}{2\sigma}\sqrt{\frac{\pi}{2}} \exp\left(
-\frac{D^2}{8\sigma^2} \right) {\rm erfc}\left(
\frac{-iD}{\sigma\sqrt{8}} \right) \ , \label{Av111}
\end{equation}
where ${\rm erfc}(...)$ is the probability integral (see, e.g.,
Ref.~\cite{AbramStig}).

With the use of (\ref{Av111}), the correlator (\ref{Av1/w}) can be
accurately computed. However, subsequent analysis can be performed
analytically only in the case of small-height roughness. With
inequality (\ref{small_height}), the averages (\ref{Av1/w2}) and
(\ref{Av1/w}) are asymptotically equal to
\begin{eqnarray}
&& \left< \frac{1}{w(t)} \right> \approx \frac{1}{D} \left[
1+4\left(\frac{\sigma}{D} \right)^2\right] \ , \label{Av1/w-appr}
\\ && \left< \frac{1}{w^2(t)} \right> \approx \frac{1}{D^2}\left[
1+12\left(\frac{\sigma}{D} \right)^2\right] \ .
\label{Av1/w2-appr}
\end{eqnarray}

The next thing to proceed is the calculation of binary correlation
functions $\left<\eta_h(x_1)\eta_h(x_2)\right>$ and
$\left<\zeta_{h\pm}(x_1)\zeta_{h\pm}^*(x_2)\right>$. In doing so,
one has to calculate the correlator
\begin{eqnarray}
{\cal L}(t_1-t_2) & =&
\left<\left[\frac{1}{w^2(t_1)}-\left<\frac{1}{w^2(t_1)}\right>\right]
\left[\frac{1}{w^2(t_2)}-\left<\frac{1}{w^2(t_2)}\right>\right]\right>
\nonumber\\& =&\left< \frac{1}{w^2(t_1)} \frac{1}{w^2(t_2} \right>
- \left< \frac{1}{w^2(t_1)} \right> \left< \frac{1}{w^2(t_2)}
\right> \ . \label{L12}
\end{eqnarray}
In the case of small-amplitude roughness, the second term in the
rhs of Eq.~(\ref{L12}) is governed by asymptotic
(\ref{Av1/w2-appr}). As for the first term, for its calculation
the method which was already applied to calculate the average
(\ref{Av1/w}) is helpful. To this end, represent the desired
correlator in the form
\begin{eqnarray}
\left< \frac{1}{w^2(t_1)} \frac{1}{w^2(t_2} \right> = \lim_{D'\to
D} \frac{\partial^2}{\partial D\,\partial D'}
\int_{}^{}\!\!\!\!\int_{0}^{\infty} dq_1 dq_2 \left< \exp \Big\{
iq_1\left[ D+\Delta\xi(t_1)+i0 \right]  - iq_2 \left[
D'+\Delta\xi(t_2)-i0 \right]\Big\} \right> \ .\qquad\quad
\label{d1d2-av}
\end{eqnarray}
Averaging of the exponent function in (\ref{d1d2-av}) yields
readily
\begin{equation}
\left< \exp \left[ iq_1\Delta\xi(t_1) - iq_2\Delta\xi(t_2)\right]
\right> = \exp\left\{ -2\left( q_1^2+q_2^2 \right) \sigma^2 + 4q_1
q_2\sigma^2 {\cal W}(t_1-t_2)\right\} \ , \label{d1d2-aux}
\end{equation}
whereupon we arrive, after some tedious manipulations, at the
integral representation
\begin{eqnarray}
\left< \frac{1}{w^2(t_1)} \frac{1}{w^2(t_2} \right> =
\frac{1}{(2\sigma)^4\sqrt{W_+ W_-}}  \int_0^{\infty} du \cos\left(
\frac{D}{\sigma\sqrt{2W_+}}u \right) \int_{\varphi(u)} ^{\infty}
dv \left( \frac{v^2}{W_-}  - \frac{u^2}{W_+} \right) \exp\left(
-\frac{u^2+v^2}{2} \right) \ .\qquad\quad \label{d1d2-fin}
\end{eqnarray}
Here, for the sake of the formula compactification, we use the
notations
\begin{equation}
\varphi(u) = u\sqrt{W_-/W_+} \ ,\qquad\qquad W_\pm = 1 \pm {\cal
W}(t_1-t_2) \ . \label{Wpm}
\end{equation}

Correlation function (\ref{d1d2-fin}) can be calculated
numerically for the arbitrary value of $\sigma/D$. But in the case
of small-height roughness the integration in (\ref{d1d2-fin}) can
be performed analytically, yielding the asymptotic result
\begin{equation}
\left< \frac{1}{w^2(t_1)} \frac{1}{w^2(t_2} \right> \approx
\frac{1}{D^4} \left[ 1 + \left(\frac{2\sigma}{D}\right)^2 \left(
5W_+ + W_- \right) \right] \ . \label{d1d2-asymp}
\end{equation}
By substituting (\ref{d1d2-asymp}) and (\ref{Av1/w2-appr}) into
(\ref{L12}) we obtain
\begin{equation}
{\cal L}(t_1-t_2)\approx \frac{16\sigma^2}{D^6} {\cal W}(t_1-t_2)
\ . \label{L12_W}
\end{equation}
After simple integration of correlator (\ref{L12_W}), resulting
from definition (\ref{Eta_hZeta_h}), we arrive at the final
expressions for the required correlators,
\begin{subequations}
\label{Eta_hZeta_h-fin}
\begin{eqnarray}
&& \left< \eta_h(x_1)\eta_h(x_2) \right> =
\frac{1}{L_f^{(h)}}F_l(x_1-x_2) \ , \label{Eta_fin} \\ &&
\left<\zeta_{h\pm}(x_1)\zeta_{h\pm}^*(x_2)\right> =
\frac{1}{L_b^{(h)}}F_l(x_1-x_2) \ . \label{Zeta_fin}
\end{eqnarray}
\end{subequations}
In Eqs.~(\ref{Eta_hZeta_h-fin}), the extinction lengths
$L_{f,b}^{(h)}$ associated with the height potential (\ref{Vh})
are given by Eqs.~(\ref{L_fhL_bh}).

%%%%%%%%%%%%%%%%%%%%%%%%%%%%%%%%%%%%%%%%%%%

\end{document}